% Modificata il 21-2-1997  
% cerchi E+F
% MNDSAMPL.TEX   last modification: 8.12.92
% A sample two column Monthly Notices article.
%
% Marginal adjustments using \pageoffset maybe required when printing
% proofs on a Laserprinter (this is usually not needed).
% Syntax: \pageoffset{ +/- hor. offset}{ +/- vert. offset}
% e.g.    \pageoffset{-3pc}{-4pc}
\input mnrass.sty
\pageoffset{-2.5pc}{0pc}

%\def\etal{{\it et al.}}
 
% \font\euler=eurm10
% \def\umu{\hbox{$\euler\mu$}} \def\upi{\hbox{$\euler\pi$}}
% Uncomment the above two lines if the Euler font is available

%\Referee   %  uncomment this for referee mode
\Autonumber  %  auto-number sections, subsections and subsubsections

% \pagerange, \pubyear and \volume are defined at the Journals office and
% not by an author.

\pagerange{000--000}
\pubyear{1996}
\volume{000}
%\umufiche{000}     % for articles with microfiche
%\authorcomment{Based on observations obtained at the European Southern
%Observatory, La Silla, Chile}  % author comment for footline

\begintopmatter  %  start the two spanning material

\title{Nuclear burning rates and population II stellar models}

\author{ E.Brocato $^{1,2}$, V.Castellani$^{2,3}$, F.L.Villante $^{4}$} 

\affiliation{$^1$Osservatorio Astronomico Collurania, I-64100 Teramo, Italy}
\affiliation{$^2$Istituto Nazionale Fisica Nucleare, LNGS, I-67100 Assergi, Italy}
\affiliation{$^3$Dipartimento di Fisica, Universit\'a degli studi di Pisa, 
I-56100 Pisa,
Italy}
\affiliation{$^4$Dipartimento di Fisica, Universit\'a degli studi de L'Aquila,
 I-67100 Italy}

\shortauthor {E.Brocato, V.Castellani, F.L.Villante}
\shorttitle{Nuclear burning rates and population II stellar models}

% \acceptedline is to be defined at the Journals office and not
% by an author.

%\acceptedline{Accepted 1994 February 30. Received 1994 February 29;
%  in original form 1994 February 28}

\abstract
\tx
We discuss theoretical predictions
concerning the evolution of globular cluster Pop.II  stars
vis-a-vis current estimates of standard errors in the
determination of nuclear burning rates. Numerical evaluations
are given for the dependence of TO and HB luminosity from
the rate of the relevant nuclear  reactions. We conclude that evolutionary
predictions  appear rather solid in this respect, with a maximum 3$\sigma$ 
error of about 1 Gyr in the evaluation of cluster ages derived from
the calibration of the difference in luminosity between HB and TO. However,
current evaluation of the original He content,
as given on the basis of the R parameter, should wait for a
much better determination of the C12($\alpha$,$\gamma$)O16
reaction before reaching a satisfactory accuracy.

\maketitle  %  finish the two spanning material

\section{Introduction}

\tx 

The theoretical evolutionary scenario for Pop.II globular cluster
stars has long been recognized as a key tool to reach a close
insight into the early evolution of the Galaxy and, in turn, on 
the history of the Universe. In recent times, theoretical constraints
concerning the age of galactic globular clusters have been 
widely debated in the literature, because of a possible
contradiction between those ages and the age of the Universe as 
derived from current estimates of the Hubble constant (see, e.g., van den 
Bergh, 1994).
In this context, of particular relevance is a recent paper by 
Chaboyer (1995) who discussed the degree of confidence to be assigned 
to theory, investigating globular cluster ages for suitable variations in
the adopted input physics. 
\par
In this paper we will follow a similar approach, discussing 
the evolutionary behaviour of low mass
Pop.II stars on the basis of current uncertainties in the nuclear burning 
rates relevant for similar structures. The consequences of an overall 
variation of H burning rates on the cluster Turn Off (TO) luminosity
has been already discussed in the above paper by Chaboyer. Here
we will enter in more details, showing that the contribution of 
pp or CNO burning  rates has to be separately taken into account, 
since an increase (or a decrease) of both pp and CNO rates has opposite
 and compensating
effects on TO luminosity. 
Moreover, we will discuss the influence of
the relevant reaction rates on current evaluation of Horizontal
Branch (HB) structures, investigating in particular HB luminosities
 and lifetimes.
\par
By inserting in such a scenario suitable values of standard 
errors for nuclear cross sections we will finally  evaluate
the corresponding errors for selected evolutionary 
features, namely the TO luminosity, the difference in
luminosity between HB and TO and the parameter R constraining
the amount (Y$_0$) of original He in cluster stars.
We will find that current uncertainties in 
the nuclear burning rates have only a minor influence on 
the theoretical predictions constraining the cluster age, but a not negligible
influence on current estimates of Y$_0$.

\section{H burning phases} 

\tx
All the evolutionary computations presented in this paper have been
produced by adopting the FRANEC stellar evolutionary code, as
already described in previous papers (see, e.g., Chieffi \& Straniero, 1989 and
reference therein).
\par
To investigate H burning evolutionary phases
a suitable set of stellar models with masses in the range 0.7 to 
0.9 M$_{\odot}$  has been followed with the standard version of
the code from the Zero Age Main Sequence (ZAMS) phase till
the onset of the He flash. We assumed 
Y=0.23  and Z=0.0001 or 0.001 as suitable original chemical composition covering the
range of metallicities found in globular cluster stars. 
The computations have been thus repeated for different assumptions about
the astrophysical factors governing  the efficiency of the 
relevant burning rates.
\par
Figure 1 compares the canonical evolutionary
track covering the H burning phases of a 0.8 M$_{\odot}$ (Z=0.0001)
model with similar tracks but 
with the pp cross section ($\sigma_{pp}$) increased by a factor 1.3 (left panel)
or the CNO cross section ($\sigma_{CNO}$) increased by
 a factor of 3 (right panel).
\par
\figure{1}{D}{110mm}{\bf Figure 1. \rm
The evolutionary path in the HR diagram of standard
stellar models (full lines) compared with similar models but with
pp burning rates increased by a factor 1.3 (left panel) or with CNO 
rates increased by a factor of 3 (right panel).}
\table{1}{S}{\bf Table 1. \rm Selected evolutionary parameters
 for a 0.8 M$_{\odot}$ star
with Z=0.0001, Y=0.23 and for the labeled assumptions on nuclear reaction 
rates} 
{\halign{%
\rm#\hfil&\hskip4pt\hfil\rm#\hfil&\hskip4pt\hfil\rm\hfil#&
\hskip4pt\hfil\rm\hfil#&\hskip4pt\hfil\rm\hfil#&\hskip4pt\hfil\rm\hfil#&
\hskip4pt\hfil\rm\hfil#\cr
Rates & logL$_{TO}$ & t$_{TO}$ & logL$_{clump}$ & logL$_{flash}$ & Mc 
& $\tau$$_{RGB}$ \cr
\noalign{\vskip 10pt}
         &       & Gyr &       &      &    & Myr \cr
Standard & 0.401 & 13.49 & 2.239 & 3.272 & 0.509 & 80.4 \cr
pp$\times$1.3 & 0.414 & 13.71 & '' & '' & ''  & '' \cr
CNO$\times$3 & 0.365 & 13.28 & 2.193 & 3.306 & 0.503 & 78.1 \cr}}

A closer insight into the role of nuclear burning rates in the
evolution is given in Table 1 where we report
selected structural parameters of the 0.8 M$_{\odot}$ model
for the labeled choices about the efficiency of reactions. Left
to right one finds the luminosity of the track TO (logL$_{TO}$), the
 evolutionary time at the TO (t$_{TO}$), the luminosity of the Red Giant (RG)
 clump (logL$_{clump}$), the 
luminosity at the onset of the He flash (logL$_{flash}$), the mass of the He
core at this time (Mc) and the Red Giant Branch (RGB) evolutionary time
($\tau$$_{RGB}$-defined as the time the star elapses between 
log(L/L$_{\odot}$)
=1.7 and the He flash).
\par
Data in table 1 show
that increased pp or CNO rates have opposite effects on the luminosity
of the evolutionary models and, in particular, on the luminosity of the
track TO. Such an occurrence can be easily understood on a quite simple
basis: let us here only recall that the track TO is marked by the
onset of CNO burning, and this onset is obviously trought forward  when
CNO rates are increased or pp rates are decreased. 
\par

As expected, one finds
that variations in pp rates do not affect at all the evolution of RG structures. 
On the contrary, increasing the CNO rates moderately decreases
the luminosity of  the RG bump marking the encounter of
the H burning shell with the chemical discontinuity produced by
the first dredge up. Moreover, one finds that  
variations modify the RG core-luminosity relation, since
the He flash is reached at a larger luminosity ($\Delta$logL$_{flash}$=0.034) 
but with a lower mass of the He core ($\Delta$Mc=-0.006). However,
 evolutionary times along the RG branch are practically unaffected.
\par
The set of evolutionary tracks has been finally used to produce
cluster isochrones under the various assumptions about nuclear
reaction rates. One finds that the sensitivity of cluster TO luminosity
to variations in burning rates is marginally
dependent on the mass of the evolving stars and thus on the assumed 
cluster age. As a whole, one can safely assume
$\Delta$logL$_{TO}$$\sim$ -0.03 as the maximum variation expected for a
decrease of pp rates by 30\% or, alternatively, for an increase
of CNO rates of 200\% .  In terms of increments one thus finds;
\par
$\Delta$ logL$_{TO} \sim {0.1} \delta{pp} -0.015 \delta{CNO}$
\par
where $\delta$pp and  $\delta$CNO represent the fractional variation
in the nuclear burning rates.

\figure{2}{D}{110mm}{\bf Figure 2. \rm
Standard ZAHB sequences (dashed-dotted lines) for Z=0.0001 
or Z=0.001  compared with ZAHB sequences computed
by (only) increasing by a factor 3 the CNO reaction rates (short dashed lines),
and by also taking into account the predicted decrease of Mc (full lines).
Dotted lines connect selected models with the same mass.}

\section{He burning phase} 

\tx
Since the luminosity of the isochrone TO is the clock marking the cluster ages,
the results discussed in the previous section will
be easily translated into theoretical uncertainties of age
estimates. However, experimental (i.e. observational) constraints on such
a luminosity can be derived from the observed TO magnitudes only 
if preliminary estimates of the cluster distance modulus (DM) are available
and suitable evaluations for the bolometric corrections are
provided. The DM is often -but not in all cases- evaluated using 
HB stars as standard (theoretical) candles; in this way one can 
directly calibrate the difference in luminosity between HB and TO 
(logL$_{HB-TO}$) 
in terms of cluster ages (see, e.g., Renzini \& Fusi Pecci 1988).
According to such a procedure, the HB luminosity behaves as a further
ingredient entering in the age determination, whose dependence on
the assumed burning rates will be investigated in this section.
\par
HB luminosity plays a role also
in the estimates of another relevant evolutionary parameter, 
the amount of original He (Y$_0$). Following a well established
procedure suggested by Iben (1968), the estimate of Y$_0$ relies
on the observational value of the parameter R, the ratio between the number
of HB stars to the number of RG stars above the HB luminosity level (see,
e.g., Renzini \& Fusi Pecci 1988). Theoretical calibration of this 
parameter, besides depending on HB luminosity, relies then on evolutionary 
predictions concerning stellar lifetimes
both in the RG and in the HB phases. Evolutionary times
along the RGB have been already discussed. In the following we will thus focus
our attention also on HB lifetimes, to reach an indication of the 
accuracy of theoretical estimates for both cluster ages and
original He content.

\figure{3}{D}{110mm}{\bf Figure 3. \rm
Dependence of HB lifetimes on the C12($\alpha,\gamma$)O16 reaction rate}

\subsection{H burning reactions}

\tx
As already discussed, pp reaction rates have no influence on the
mass of the He core at the onset of the He flash (Mc). Nor do variations 
in these rates affect the distribution of chemical elements
throughout the structure of a new-born HB star. Bearing in mind 
that H shell burning in HB stars is dominated by CNO reactions, one 
easily foresees that HB evolution should be largely unaffected by 
changes in pp rates. This prediction is indeed confirmed by 
numerical experiments. By decreasing the pp rate by 30\% one finds
that the Zero Age Horizontal Branch (ZAHB) locus remains unchanged,
 with only a slight increase of the 
masses located at a given effective temperature. As an example,
 with the quoted decreased
rate, the mass of the model at logTe=3.85 is M=0.802 M$_{\odot}$ against M=0.796  
M$_{\odot}$ for the canonical case.
At the same time He burning evolution of the models does not show
 relevant variations.
\par
CNO burning rates deserve more attention, since they affect
both Mc and the energy sources in HB stars. The rather complex behaviour
of ZAHB models with CNO 
rates increased by a factor of 3 is shown in Figure 2, where
canonical ZAHBs for Z=0.0001 or Z=0.001 are compared with similar loci 
but computed i)
by increasing only CNO rates and, ii) by considering also the predicted
decrease of Mc. The increase of the
CNO rates drives a corresponding increase of the ZAHB luminosity, whereas
the decrease of Mc works in the opposite direction. However, Figure 
2 shows the unexpected result that the net affect of
these two contributions to ZAHB luminosity substantially depends on 
the assumed metallicity.
In more detail, at logTe=3.85, one finds that the luminosity of
the ZAHB {\it decreases} by about $\Delta$logL=0.017 when Z=0.0001, whereas
it {\it increases} by about the same amount if Z=0.001. The same Figure
2 shows that such a behavior follows the larger variation in luminosity
induced by the increased CNO rates in the Z=0.001 case, due to the greater
 efficiency of CNO burning in similar moderately
metal rich structures. According to such evidence, one finds that
when Z=0.0001 the decrease
of Mc succeeds in pushing the final ZAHB luminosity below the original 
value, whereas for Z= 0.001 the increase in luminosity prevails 
over the effect of Mc.

\table{2}{S}{\bf Table 2. \rm The expected 
variation of the labeled evolutionary parameters
when the various nuclear cross section are increased by a factor of 2.} 
{\halign{%
\rm#\hfil&\hskip10pt\hfil\rm#\hfil&\hskip10pt\hfil\rm\hfil#&
\hskip10pt\hfil\rm\hfil#&\hskip10pt\hfil\rm\hfil#\cr
\noalign{\vskip 10pt}
& $\Delta$$pp$ & $\Delta$$CNO$ & $\Delta$$3$$\alpha$ &
$\Delta$$C$$^{12}$,$\alpha$ \cr
$\Delta$logL$_{TO}$ & 0.1 & -0.015 & - & - \cr
$\Delta$logL$_{HB}$ & - & $\pm{0.009}$ & -0.040 & - \cr
$\Delta$$\tau_{HB}/\tau$ & - & - & -0.02 & 0.10 \cr}}

\subsection{He burning reaction}

\tx
The 3$\alpha$ reaction starts affecting HB structures
from the final phase of RG evolution, governing the onset of the He
flash and, in turn, the mass of the He core in a new born ZAHB
star. Numerical experiments show that decreasing the
3$\alpha$ rate by a factor 0.7  increases Mc by $\Delta$Mc=
0.003. According to the pioneering paper by Sweigart \& Gross (1976)
we know that HB luminosity depends on Mc according to the
relation $\Delta$logL$_{HB}$=3.4$\Delta$ Mc. Consequently the quoted 
variation of
Mc should, alone, induce an increase of 0.010 in the ZAHB
luminosity. Numerical experiments disclose that  
such a prediction is only slightly
increased by the contribution given by decreased 3$\alpha$ rates. As a result,
one finds that for the given variation of 3$\alpha$ rates ZAHB
luminosity will increase by about $\Delta$logL=0.012.
In the meantime, the lifetime $\tau$$_{HB}$ 
for the entire phase of central He burning increases by about 0.7\% .
\par
To study the role of the C12($\alpha$,$\gamma$)O16 reaction
we computed selected HB models by increasing that rate ($\sigma_{C12}$) by
selected factors. As expected, HB lifetime
increases since the completion of the chain of reactions is favored
and more energy is obtained from the He burning. Figure 3 shows that
HB lifetimes do not depend linearly on $\sigma_{C12}$. This could be expected
since HB lifetimes
cannot be affected by $\sigma_{C12}$ in the limits where all or no carbon 
is converted in oxygen. Figure 3 however shows that in the range
$\sigma_{C12}/\sigma_{C12}^{sta}\le2$ one can adopt a linear approximation
given by $\Delta\tau_{HB}/\tau_{HB}\sim0.10\Delta\sigma_{C12}/\sigma_{C12}
^{sta}$ . Note that one finds HB lifetimes increased by values
 larger than predicted by previous estimates reported in Renzini \& Fusi 
Pecci (1988).
\par
Table 2 summarizes the results of all the numerical experiments,
giving the dependence of TO luminosity, HB luminosity and HB lifetimes
on changes in the rates of the relevant nuclear reactions.

\table{3}{S}{\bf Table 3. \rm The 3$\sigma$ errors for the labeled reactions and 
the corresponding uncertainties in logL$_{TO}$, logL$_{HB}$ and logL$_{HB-TO}$.
Last row gives the total 3$\sigma$ error
obtained by quadratically compounding the pp, CNO and $3\alpha$ uncertainties} 
{\halign{%
\rm#\hfil&\hskip3pt\hfil\rm#\hfil&\hskip3pt\hfil\rm\hfil#&
\hskip3pt\hfil\rm\hfil#&\hskip3pt\hfil\rm#\hfil&\hskip3pt\hfil\rm#\hfil\cr
Rates & Errors & $\Delta$logL$_{TO}$ & $\Delta$logL$_{HB}$ &
$\Delta$logL$_{HB-TO}$ &  \cr
\noalign{\vskip 10pt}
     &      &       &   &  Z=10$^{-4}$ & Z=10$^{-3}$ \cr 
pp & 5\% & 0.005 & - & 0.005 & 0.005 \cr
CNO & 51\% & 0.008 & 0.005 & 0.003 & 0.013 \cr
3$\alpha$ & 45\% & - & 0.018 & 0.018 & 0.018 \cr
3$\sigma$$_{tot}$ &   & 0.009 & 0.019 & 0.019 & 0.023 \cr}}

\section{Standard errors}

\tx
We are now in the position to link  numerical experiments 
discussed in previous sections with current estimates for standard
errors in burning rates in order to estimates standard
errors in the predicted evolutionary features. 
\par
For pp reactions, from the recent European cooperation on
nuclear data (NACRE97) one derives
 S(0) = 3.94 (1 $^{+0.075}_{-0.025}) 10^{-25}$
Mev barns (Degl'Innocenti, private communication). As a consequence
we assume in the following a standard error of 5\% (at 3 $\sigma$).
Concerning CNO rates, Rolf \& Rodney (1988) give for the the main reaction
N14(p,$\gamma$)O15 a range of value as large as S(0)= 2-10 kev barns.
However, this range appears greatly reduced in Schr\"oder et al. (1987),
which gives a 1 $\sigma$ error of about 17\%. In the following
 we will adopt this last result.
According to Rolf \& Rodney (1988) the error in the 3$\alpha$
reaction can be again estimated to be of the order of 15\%. However
the uncertainty in the C12($\alpha$,$\gamma$)O16 reaction rate is much worse.
According to Caughlan \& Fowler (1988) reaction rates usually adopted
in current evolutionary computations could be wrong by
a factor of 2 up or down. In a more recent determination 
Buchmann (1996) limits S(300) to the range 62-270 keV barns.
 \par
By relying on the above estimates it is 
now possible to carry out the exercise of deriving expected standard errors on
the luminosities of TO and HB. The results of this exercise are reported 
in Table 3, where we list for each given reaction 
the assumed 3$\sigma$ errors and the corresponding expected variation 
in evolutionary parameters.
\par
Concerning the calibration of logL$_{TO}$ in
terms of cluster age, since dlogL$_{TO}$$\sim$dlogt$_{9}$ (see, e.g., 
Castellani \& Degl'Innocenti 1995), for cluster 
age of 14 billion years we obtain a standard error  
of the order of 0.3 Gyr, in reasonable  agreement 
with the values given by Chaboyer (1995).
 From the same table one finds that the HB luminosity is substantially
affected only by CNO and 3$\alpha$ rates.
\par
In the last two columns of
table 3 we report the uncertainties of the difference between the TO and the HB
luminosities. In the evaluation of this quantity it is 
 necessary, in principle, to consider separately the two cases Z=0.0001 and
 Z=0.001, due to the opposite behaviour of the HB luminosities discussed in 
 section 3.1. However the general result is that the uncertainty in 
 logL$_{HB-TO}$ is,
 at most, of order 0.023 .As a results, when calibrating ages through the
 difference in magnitude between HB and TO, the error is of the order 
 of magnitude of 0.8 Gyr.
\par
Data in Table 2 allow us finally to  investigate the influence of
"nuclear" errors on current estimates of the amount of the original
He, as based on the calibration of the R parameter in terms of Y$_0$.
By recalling that R is theoretically defined as the ratio of He 
burning lifetime to RG lifetime  at luminosity larger than
the HB luminosity, one finds that the relevant reactions are
CNO and 3$\alpha$ which affect the HB luminosity only, and C12+$\alpha$
which governs HB lifetime. 
\par
By relying on available estimates of the lifetime along
the RG branch (see, e.g., Bono et al. 1995) one finds that the
uncertainty in the HB luminosity level translates into an uncertainty
 of 3.5\% in the
evolutionary time of an RG above this level, and a corresponding
error in the parameter R. On the basis of theoretical calibration
of such a parameter (see again Bono et al. 1995) one
concludes that the contribution to the 
Y$_0$ error is rather negligible, less
 than 0.01.
% E+F Se abbiamo fatto bene i calcoli questo valore e' 0.01 o
% 0.006 a seconda che la derivata dY/dR venga calcolata tra Y=0.20 e 0.23
% oppure tra Y=0.23 e 0.27 (tab.5 - Bono et al.) 
\par
Since current calibrations of R are based on
 Caughlan et al. (1985) reaction rates an error of the order of the factor
 2 has been assumed to investigate the relevance of the
 C12($\alpha$,$\gamma$)O16 reaction rates uncertainties.
 With such an error the C12($\alpha$,$\gamma$)O16
reaction affects the He burning evolutionary time by about 10\%.
Again on the basis of current R calibration, it turns out
that the theoretical calibration of He is affected 
by an error of order
$\Delta Y_0\sim 0.02$ which appears now of some relevance.
As an example, increasing $\sigma_{C12}$ by
a factor two would move Y$_0$ from Y$_0$=0.25
 to Y$_0$=0.23. 
% E+F Come sopra: 0.02 se la derivata e' calcolata tra Y=0.20 e 0.23
% oppure 0.01 se la derivata e' calcolata tra Y=0.23 e 0.27 
% E+F Rispetto alla versione precedente e' diminuito l'errore su Thb
% dovuto alla C12 (come abbiamo discusso al telefono. Dal 15% all' 8%).
% Comunque sarebbe opportuno controllare gli errori su Yo.
% E+F Gli errori su Yo dovuti a C12 e 3alpha forse dovrebbero essere composti.
% In tal caso e' giusto comporli quadraticamente oppure semplicemente
% sommarli?

\section{Conclusions} 

\tx

In this paper we have discussed the reliability of theoretical prescriptions
concerning the evolution of globular cluster Pop.II  stars
vis-a-vis current estimates for standard errors in the
determination of nuclear burning rates. We computed stellar evolutionary
models adopting different assumptions for the reaction
rates involved during the H and He burning phases, evaluating the dependence
of selected evolutionary features on the nuclear reaction rates.
 The main results can
be summarized as follow:
\medskip
\noindent
i) The uncertainties in pp rates sligthly affect only the TO luminosity;
\smallskip\noindent
ii) The CNO rates uncertainties may introduce a small variation in the TO and
HB luminosities. Due to the different efficiency of CNO burning, the effect of 
a given variation of CNO on HB luminosity depends on the assumed 
metallicity;
\smallskip\noindent
iii) An increase (or a decrease) in the pp rate has the opposite effect on TO
luminosities of an increase (or a decrease) in the CNO rates, almost 
removing any
sizeable variation in logL$_{TO}$;
\smallskip\noindent
iv) The uncertainty of the 3$\alpha$ rates affects the HB luminosity but has a 
negligible effect on the HB evolutionary time;
\smallskip\noindent
v)  The uncertainty of the 
C12($\alpha$,$\gamma$)O16 rate  seems to introduce a non-negligible
 uncertainty in the HB lifetime.
\medskip

On this basis we investigated the relevance of the current nuclear rection rates    
uncertainties to theoretical calibrations concerning cluster ages and
the original He content Y$_{0}$. We conclude that cluster ages derived from the
calibration of the difference in luminosities between HB and TO are affected by
 maximum 3$\sigma$ error of about 1 Gyr. However,
it appears that the theoretical calibration of the parameter R
 should wait for better determination of the C12($\alpha$,$\gamma$)O16
reaction before reaching a satisfactory accuracy.
\medskip

We would like to thank S.Degl'Innocenti for useful discussions and comments.

\section*{References}

\bibitem {Bono G., Castellani V., Degl'Innocenti S, Pulone L., 1995,
 A\&A, 297, 115}
\bibitem {Buchmann L., 1996, ApJ, 468, L127}
\bibitem {Castellani V. \& Degl'Innocenti S., 1995, A\&A, 298, 827}
\bibitem {Caughlan G.R., Fowler W.A., Harris M.J. \& Zimmerman B.A., 1985,
 Atomic Data and Nuclear Data Tables, 32, 197} 
\bibitem {Caughlan G.R. \& Fowler W.A., 1988, Atomic 
 Data and Nuclear Data Tables, 40, 283}
\bibitem {Chieffi S. \& Straniero O., 1989, ApJS, 71, 47}
\bibitem {Chaboyer B., 1995, ApJ, 444,L9}
\bibitem {Iben I. Jr, 1968, Nature, 220, 143}
\bibitem {NACRE97-The European NACRE Compilation, to be published in At.
 Data. Nucl. Data Tables (1997)} 
\bibitem {Renzini A. \& Fusi Pecci F., 1988, ARA\&A, 199}
\bibitem {Rolfs C.E. \& Rodney W.S. 1988, in "Cauldrons in the Cosmos",
     The University of Chicago Press}
\bibitem {Schr\"oder et al., 1987, Nuclear Physics A 467, 240}
\bibitem {Sweigart A.V.  \& Gross P.G., 1976, ApJS, 32, 367}
\bibitem {van den Berg S., 1994, PASP, 106, 1113} 

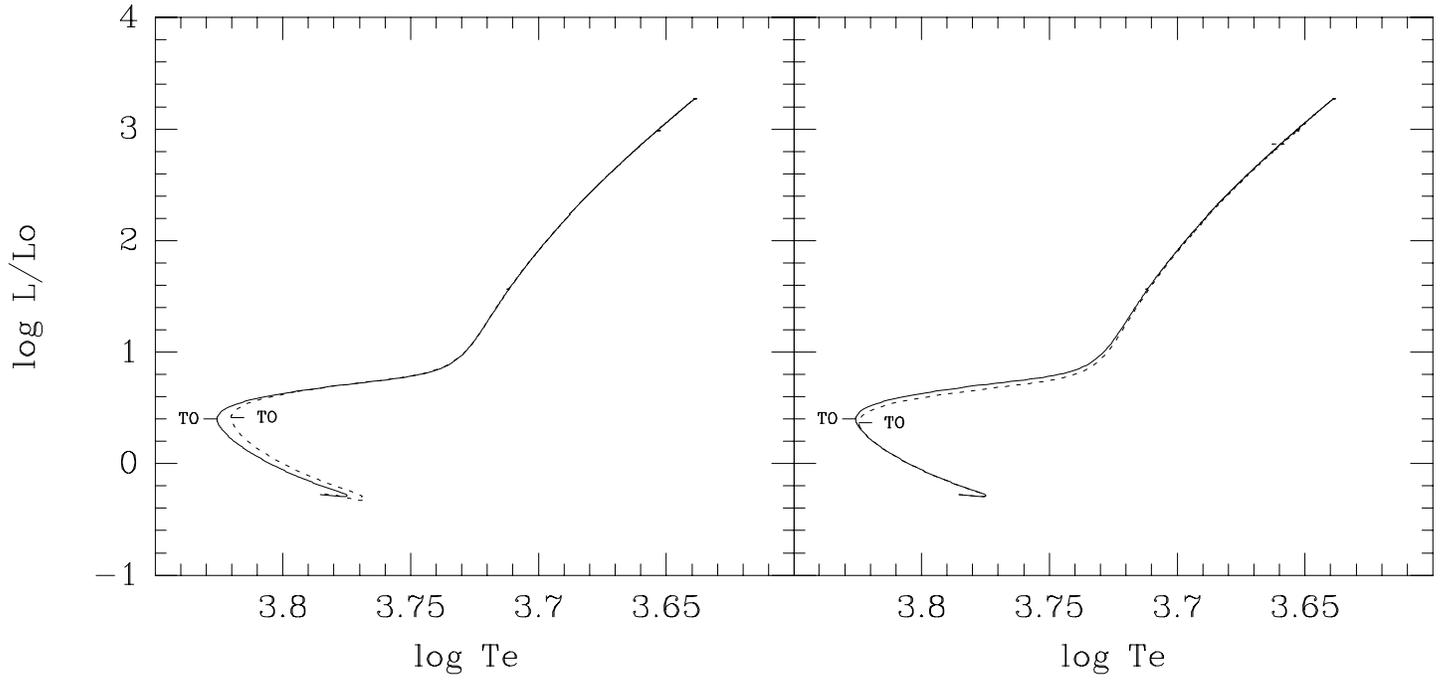
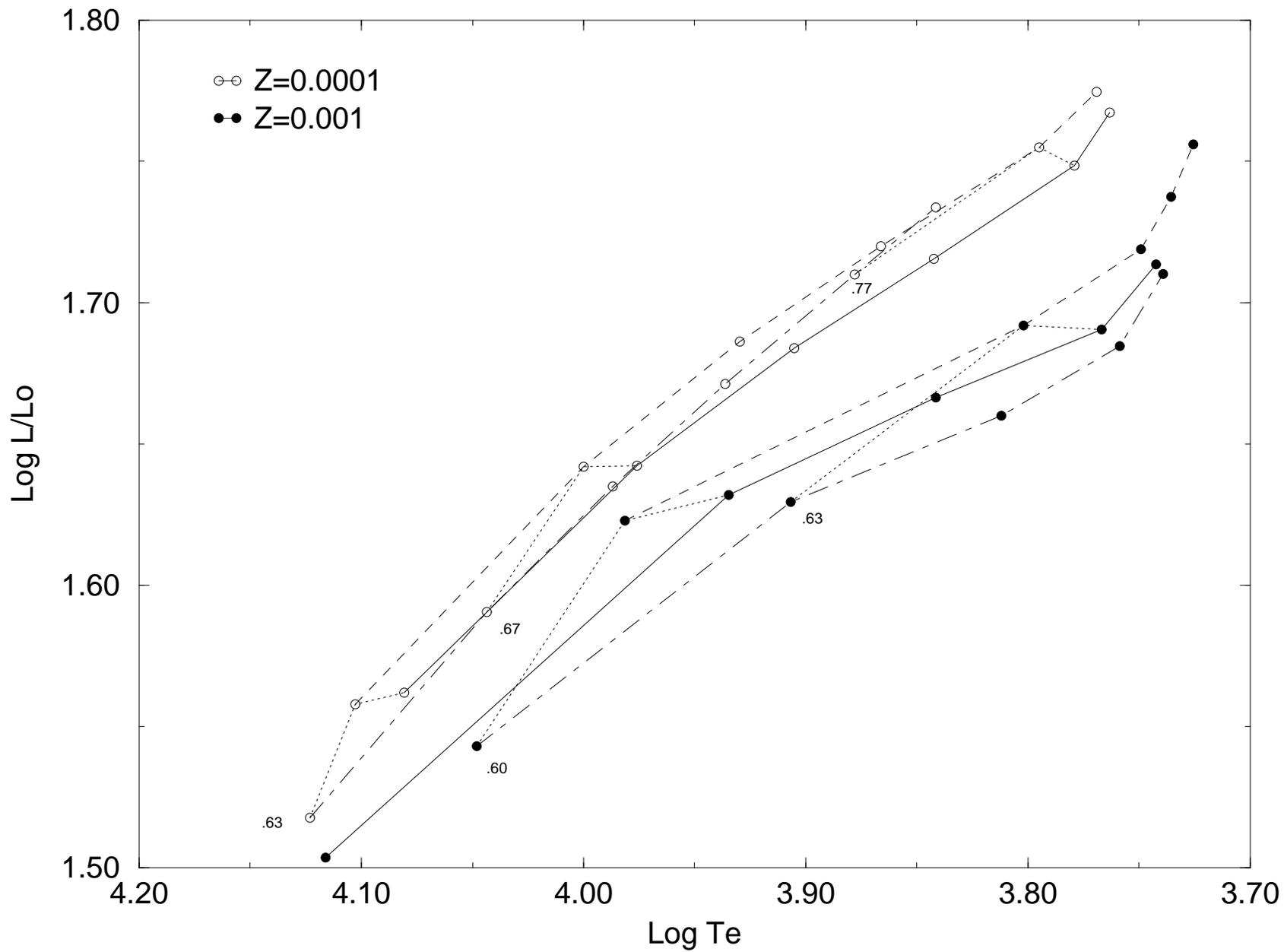
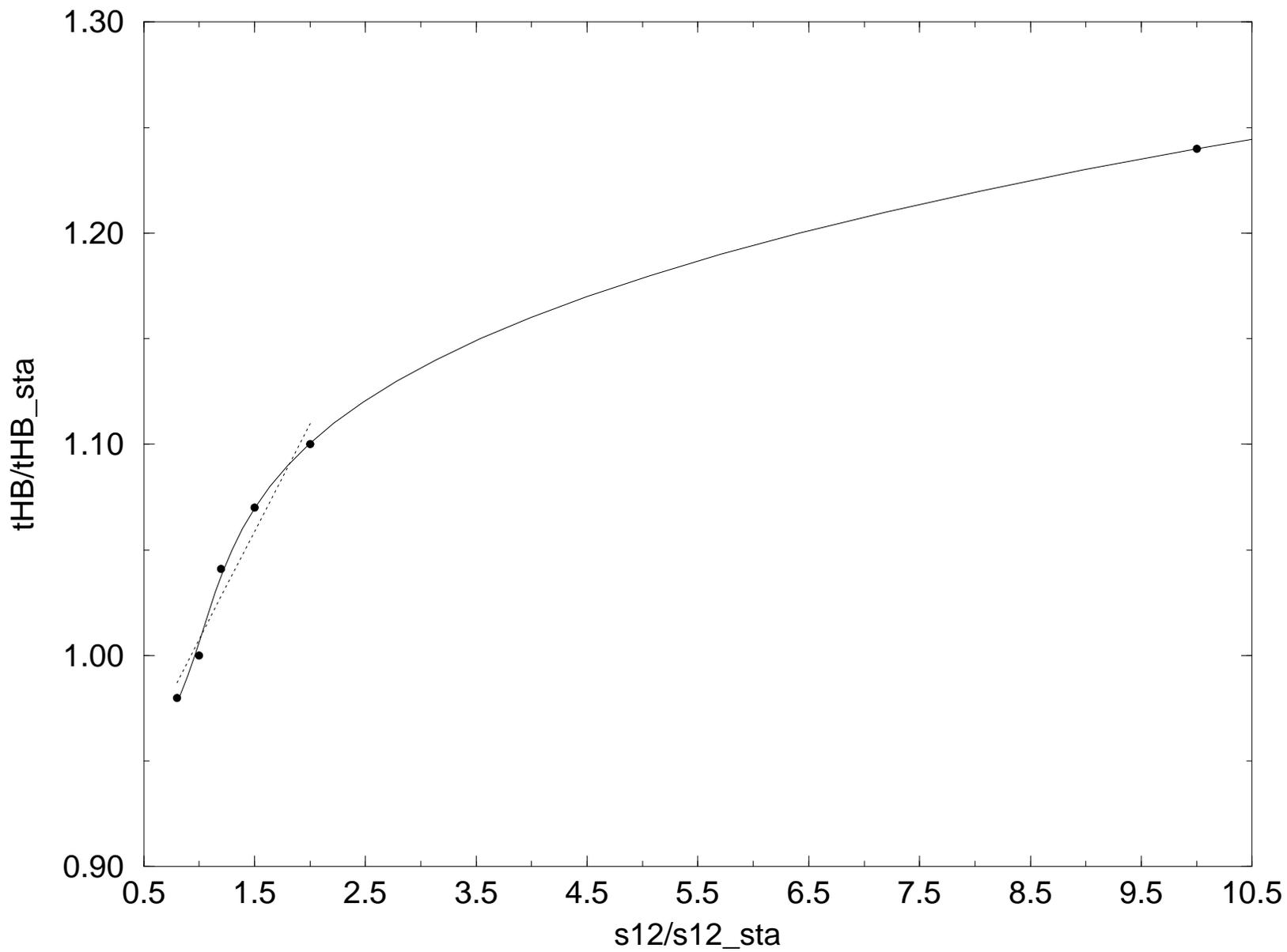
\bye